\documentclass[prl,twocolumn,amsmath,amssymb,floatfix,showpacs]{revtex4}

\usepackage{graphicx}
\usepackage{dcolumn} 
\usepackage{bm}      
\usepackage{color}

\begin{document}

\title{Persistent cyclonic structures in self-similar turbulent flows}
\author{P.D. Mininni$^{1,2}$ and A. Pouquet$^2$}
\affiliation{$^1$ Departamento de F\'\i sica, 
     Facultad de Ciencias Exactas y Naturales, Universidad de Buenos Aires, 
     Ciudad Universitaria 1428 Buenos Aires, Argentina. \\
             $^2$ NCAR, P.O. Box 3000 Boulder CO 80307-3000, U.S.A.}
\date{\today}

\begin{abstract}
Invariance properties of a physical system govern its behavior: energy conservation in turbulence drives a wide distribution of energy among modes, as observed in geophysics, astrophysics and engineering. In hydrodynamic turbulence, the role of helicity (which measures departures from mirror symmetry) remains unclear since it does not alter this distribution. However, the interplay of rotation and helicity leads to significant differences. Using numerical simulations we show
the occurence of long-lived laminar cyclonic vortices together with turbulent vortices, reminiscent of recent tornado observations. Furthermore, the small scales are completely self-similar with no deviations from Gaussianity. This result points to the discovery of a small parameter in rotating helical turbulence.
\end{abstract}

\maketitle

Self-similar behavior in turbulence is viewed as a large-scale phenomenon, recently linked to conformal invariance (e.g., local scale invariance through transformations that preserve angles but not distances) when examining the scaling properties of vortex lines \cite{bernard06} for two-dimensional turbulence \cite{kraichnan_montgo}, and surface quasi-geostrophic turbulence \cite{bernard07} for a rotating stably stratified layer  \cite{pierrehumbert}. The link found to percolation theory allows for the analytical determination of scaling exponents such as the fractal dimension of vortex clusters, although the consequences for statistical measures of turbulence, e.g., through scaling laws for correlation functions, has not been clarified yet. However, the three-dimensional case is known to be much more complex. In three dimensional turbulence, the flow is not scale invariant and the knowledge of one exponent does not allow the prediction of the exponents for all orders. The search for self-similar quantities in three-dimensional turbulence is a long-standing problem; it would relate its study with critical phenomena and the out-of-equilibrium statistics of systems with a large number of modes, and it would allow the use of tools (such as the renormalization group \cite{ma_mazenko}) from quantum field theory and statistical mechanics to further our understanding of such flows. 

One of the most important and useful principles of physics is that of conservation laws linked, through the theorem of Emma Noether, to invariance properties of the underlying equations. Conservation of energy is invoked when explaining the observation of a wide range of excited scales in a turbulent flow: the nonlinear coupling due to advection leads to the feeding of modes at all the scales available to the system  \cite{frisch_book}; this constant flux direct cascade of energy to small scales is arrested by dissipative processes. The conservation of helicity $H=\left<{\bf u} \cdot \mbox{\boldmath $\omega$}\right>$, the correlation between the velocity ${\bf u}$ and the vorticity $\mbox{\boldmath $\omega$}=\nabla \times {\bf u}$, was discovered much later \cite{moffatt}, although it appears rather clearly when formulating the Euler equations governing the dynamics of the ideal incompressible fluid in terms of the so-called Lamb vector ${\bf u} \times \mbox{\boldmath $\omega$}$. Helicity is not positive definite, unlike energy; it is a topological invariant, representing the degree of knottedness of vortex lines \cite{moffatt}, and it is a pseudo-scalar (the symmetry group related to its conservation is discussed in \cite{yahalom}). Helicity in turbulent flows can lead to drag reduction, and to better mixing of chemical components \cite{chemical}. 

\begin{figure}
\includegraphics[width=8cm]{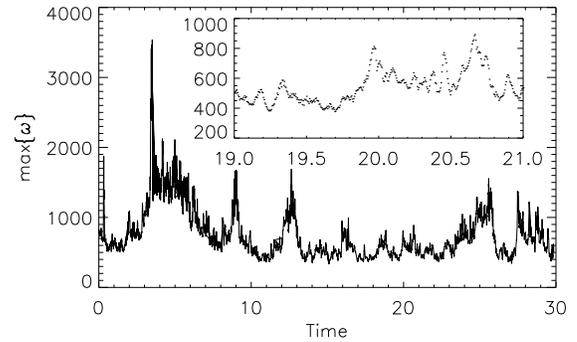}
\caption{Temporal evolution of the maximum vorticity in the flow, with $\omega_{rms}\sim 20$. The insert gives a short-time blow-up in which the inertial wave period $1/\Omega$ with $\Omega=9$ is slightly visible; note the stong local bursts in $\omega_{max}$.}
\label{fig:max} \end{figure}

Turbulent flows are known to develop helical structures which are persistent since their associated non-linear advection is weak \cite{pelz}; thus, the evolution of these structures takes place on the slow dissipative time scale. In the atmosphere, the persistence of helical flows was invoked to explain structures encountered in supercell storms that can give rise to tornadoes \cite{davies_jones}. Even though small-scale structures are found to be strongly helical, the helicity of a flow does not seem to alter its dynamics: indeed, numerical evidence stemming from direct numerical simulations of incompressible, isotropic, and homogeneous turbulence indicate (both for weak or strong global helicity) that the distribution of energy among scales follows a power-law \cite{frisch_book} which, expressed in terms of correlation functions or of structure functions of second order, reads $S_2(\ell)\sim \ell^{2/3}$, with $S_p(\ell)= \left<[u_L({\bf r}+\mbox{\boldmath $\ell$})-u_L({\bf r})]^p\right>$ the $p$th-order longitudinal structure function on a distance $\ell$, $u_L$ being the projection of the velocity field along the vector $\mbox{\boldmath $\ell$}$. Similarly, it was shown using the renormalization group \cite{pouquet_helical} that the helical contribution to eddy viscosity is sub-dominant.

The question we now address is the interplay of helicity with rotation, included in the incompressible Navier-Stokes equations through the Coriolis force:
\begin{equation}
\frac{\partial {\bf u}}{\partial t} + \mbox{\boldmath $\omega$} \times
    {\bf u} + 2 \mbox{\boldmath $\Omega$} \times {\bf u}  =
    - \nabla {\cal P} + \nu \nabla^2 {\bf u} + {\bf F} ,
\label{eq:momentum} \end{equation}
with $\nabla \cdot {\bf u} =0$; we choose the rotation axis to be in the $z$ direction: $\mbox{\boldmath $\Omega$} = \Omega \hat{z}$, with $\Omega$ the rotation frequency; ${\cal P}$ is the total pressure divided by the (constant) mass density and modified by the centrifugal term, $\nu$ is the kinematic viscosity and ${\bf F}$ is an external mechanical force that drives the turbulence (mimicking for example a convective input of energy); it is given by the ABC flow \cite{ABC} (a Beltrami flow with $\mbox{\boldmath $\omega$}$ proportional to ${\bf v}$) with $L_F=2\pi/k_F$ the characteristic scale of the forcing, of amplitude $F_0$:
\begin{eqnarray}
{\bf F} &=& F_0 \left\{ \left[B \cos(k_F y) + 
    C \sin(k_F z) \right] \hat{x} + \right. {} \nonumber \\
&& {} + \left[C \cos(k_F z) + A \sin(k_F x) \right] \hat{y} + 
   {} \nonumber \\
&& {} + \left. \left[A \cos(k_F x) + B \sin(k_F y) \right] 
   \hat{z} \right\}.
\label{eq:ABC}
\end{eqnarray}

\begin{figure}
\includegraphics[width=8cm]{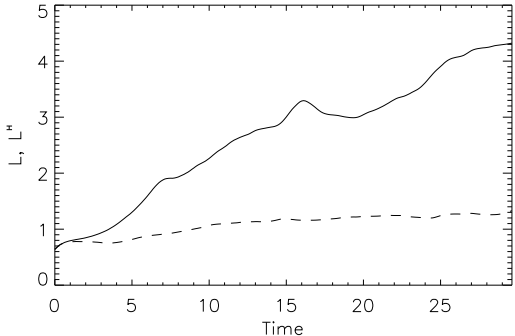}
\caption{Characteristic (integral) length scales for the energy (solid line, $L$) and the helicity (dash line, $L^H$). The significant growth of $L$ confirms the occurrence of a build-up of energy at large scale, in what is called an inverse cascade.}
\label{fig:isotropy} \end{figure}

\begin{figure} \includegraphics[width=8.5cm]{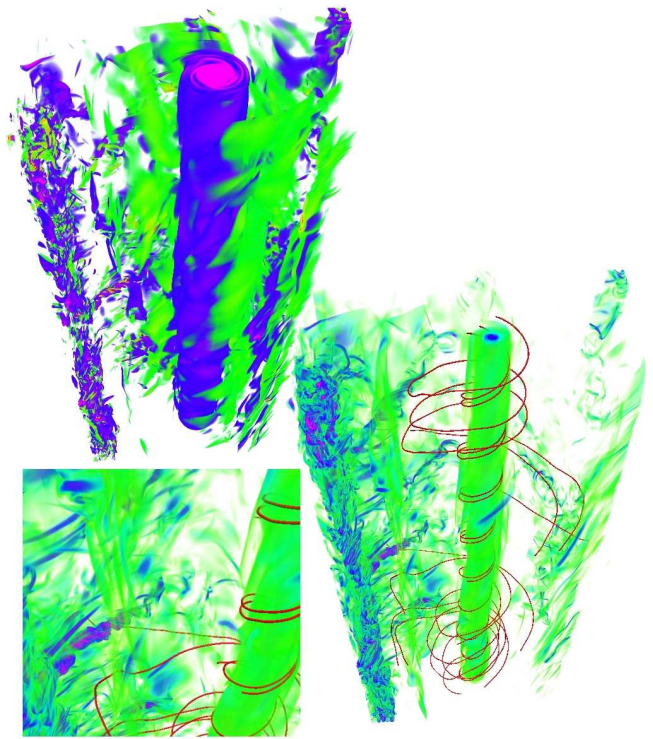} 
\caption{(Color online) Top left: visualization of helicity density at $t\approx 21$; light (green) is negative and dark (blue-magenta) is positive. Right: vorticity intensity in the same region, with fluid trajectories in red. Note the co-location of laminar structures, with smooth paths, and a tangle of vortex filaments with more complex paths. Bottom left: A zoom on the vorticity magnitude, illustrating the co-existence of both structures at very disparate scales. Small-scale vortex filaments to the left entangle to form filaments at larger scales.}
\label{fig:PVR_W+lines} \end{figure}

The above equations are integrated with a pseudo-spectral code and periodic boundary conditions using a second-order Runge-Kutta temporal scheme. Computations were performed on a grid of $1536^3$ points. The flow is first led to establish a statistically steady state with $\Omega=0.06$, i.e. in the near absence of rotation, a phase taking  roughly ten turn-over times; with $k_F=7$, the rms velocity is $U\approx 1$ and the Reynolds number $R_V=UL_F/\nu \approx 5600$ with the choice of $\nu=1.6\times 10^{-4}$. Then, at a time labeled $t=0$ in the following, the rotation is set to $\Omega=9$, corresponding to a Rossby number 
 $Ro= U/(2\Omega L_F)\approx 0.06$. The time step is $\Delta t=2.5\times10^{-4}$. The computation is then performed for 30 turn-over times $\tau_{NL}=L_F/U$, corresponding to 180 in units of a large-scale inertial wave period.

An example of the temporal evolution of the flow is given in Fig. \ref{fig:max}, showing the maximum of vorticity, the rms value of vorticity being $\omega_{rms}\approx 20$; a blow-up given in the inset shows a slight signature of  the period  of inertial waves, with of the order of nine crests per unit time; also note that the vorticity in the flow is intermittent, both in its time evolution as in its distribution in space. Fig. \ref{fig:isotropy} gives as a function of time the integral scale for the velocity (solid line) and for the helicity (dash line): the monotonous increase in the characteristic scale of the energy is a manifestation of a self-organization process of the flow at large scales, into structures that contain most of the energy. This constant flux transfer of energy towards large scales is known as an inverse cascade. On the other hand, the characteristic scale of the helicity remains approximately constant, a manifestation of helicity being transfered towards small scales.
 
Two results are striking; they concern the structures that develop in the flow and the ensuing statistics. The spatial helicity distribution and vorticity magnitude in a box of size $0.25\times0.5\times1$ the total fluid volume, and with the vertical direction upward, is given in Fig. \ref{fig:PVR_W+lines}: we observe the juxtaposition in space, for the same physical variables, of columnar laminar structures, and of strong vortex filaments in a tangled network. Within the column there is a strong alignment of ${\bf v}$ and $\mbox{\boldmath $\omega$}$ with a resulting strong local helicity density. As a result, particle trajectories (in red) are helical and upward: these are cyclonic events that are persistent in time because of the lack of nonlinearities in their vicinity. At any given time, two or three such columns can be easily identified in the entire domain. The structures are long-lived, and were tracked for over ten turnover times without being substantially deformed or destroyed by the surrounding turbulent flow. On the other hand, the tangle of vortex filaments right next to these laminar structures has complex particle trajectories, lives of the order of one eddy turn-over time and corresponds to a classical turbulent fluid albeit with some perceptible vertical organization (which is also found in the case of non-helical rotating fluids \cite{mininni_rot_TG}).

Overall, the vorticity is strongest in the tangle of vortex filaments, whereas the vertical velocity is coherent and strongest in the laminar columns (not shown). The core of the column is surrounded by a calm region with weak vorticity (note the emptiness of vortical structures in the surroundings of the column) which acts as a transition region between the laminar and the turbulent flow. The origin of these stable structures can be identified  by integrating backward in time the particle trajectories: they correspond to regions of large helicity where the columns form. Such a laminar organization of the velocity field is not observed in helical isotropic and homogeneous turbulence, nor is it observed in rotating flows without helicity. The interplay between rotation (which breaks the mirror-symmetry in the evolution equations) and helicity (which quantifies departures from mirror-symmetry of the flow) is the driving agent for the formation of strong localized and persistent columnar structures, even though helicity is mostly transfered towards smaller scales \cite{mininni_rot_hel} and is itself strongly intermittent (see below, Fig. \ref{fig:pdf}). The tangle of vortex filaments surrounding a laminar structure, together with more complex, larger and spiraling features, is reminiscent of observations of multiple core vortex tornadoes \cite{DOW}.

\begin{figure}
\includegraphics[width=8cm]{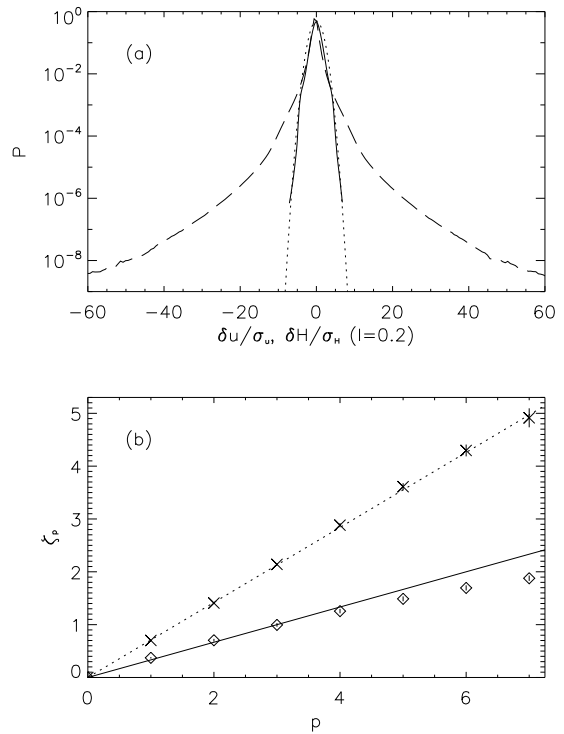}
\caption{(a) Probability density function of increments for the velocity (solid line) and helicity (long dash), normalized by their variances, averaged over ten turnover times, and taken on a distance $\ell=0.2$ (roughly 22\% of the forcing scale $L_F$, i.e., in the inertial range of the direct cascade). The velocity is close to a Gaussian (dots), whereas the helicity displays strong wings, characteristic of strong localized events.
(b) Scaling exponents of the velocity structure functions, with error bars. Diamonds are for the flow with $\Omega = 0.06$, displaying a classical three-dimensional intermittent fluid scaling. A straight line with $\zeta_2 = 2/3$ (solid) is shown as a reference; the deviation from the straight line indicates a departure from self-similarity. Crosses are for the flow with $\Omega = 9$, averaged over twelve directions in the plane perpendicular to the rotation axis, and over ten snapshots from $t=20$ to $t=30$. The dotted straight line indicates self-similar scaling with $\zeta_2=1.42$, close to the steepest scaling that can be obtained in such a flow, as found in \cite{mininni_rot_hel} using a simple argument based on a direct helicity cascade mediated by inertial waves.}
\label{fig:pdf} \end{figure}

The coexistence of the laminar columns with the turbulent flow has a strong impact on the statistics of the velocity field. This can be measured by studying the histograms of velocity and helicity differences over a distance $\ell$ chosen to be in the small scale inertial range, as well as by the computation of the exponents $\zeta_p$ of the velocity field structure functions $S_p(\ell)\sim \ell^{\zeta_p}$ previously defined. The increments were taken in twelve directions perpendicular to the axis of rotation, and averaged over volume and over ten snapshots of the fields, spanning ten turnover times. This gives a total of over $4.3\times10^{11}$ data points for each value of the scalar increment $\ell$.

The resulting histograms (Fig. \ref{fig:pdf}) are strikingly different: the velocity increments are close to Gaussian, whereas for the helicity increments strong wings obtain, characteristic of the presence of intense events and intermittency. Scaling exponents of the velocity up to $p=7$ are shown in Fig. \ref{fig:pdf}(b). In the run with $\Omega = 0.06$, the exponents are those observed in turbulence without rotation, with for example $\zeta_3=1$, a trace of the conservation of energy in the absence of dissipation in those variables. The deviation of the exponents from a straight line indicates the break down of scale invariance in the flow, as is well known for three dimensional turbulence: strong events are more probable at small scales. In the run with $\Omega = 9$, as times grows the system evolves toward a different regime: that of self-similarity where the knowledge of one exponent (say, that of second-order correlation functions)  gives scaling for all higher-order structure functions. Henceforth, the scaling of strong gradients is completely determined if one of the exponents is known; the dotted line ($\zeta_p\sim 0.71p$) represents the data best, in agreement with predictions made in \cite{mininni_rot_hel} for $\zeta_2$.

This is a remarkable property, to our knowledge never before observed in small-scale three dimensional turbulence with a direct cascade of energy. While the energy has organized in fully helical updrafts at large scales, the helicity itself concentrates in the cores of these structures as well as in the small-scale tangle of vortex filaments. The identification of  self-similarity in turbulence with both helicity and rotation (conditions that are relevant in many atmospheric flows) would allow us to relate the dynamics of such  three-dimensional flows to the advances made in two-dimensional turbulence and critical phenomena in general \cite{bernard06}. However, in order to be able to use renormalization group techniques \cite{ma_mazenko,pouquet_helical} and other tools from statistical mechanics, a small parameter needs to be identified, besides the Rossby number that governs the energetic exchanges between turbulent eddies and waves when dealing, e.g., with the weak turbulence regime \cite{cambon_NJP}. Indeed, the smallness of the ratio of the inertial wave period to the eddy turn-over time has already been used to derive integro-differential equations in terms of energy and helicity spectra \cite{Galtier03} in the context of weak turbulence. However, these solutions are not observed in our study or in many atmospheric flows for at least three reasons: first, the Rossby number in the atmosphere and in our simulations is moderate and far from the limit considered in the theory; second, the theory is non-uniform in scale and the weak turbulence limit breaks down; third, in the case of rotation, the inverse cascade of energy is not present at lowest order in the theory and thus the solution selected by this approach is one of an energy cascade to small scales, whereas our numerical data indicates that this cascade is sub-dominant to the helicity cascade \cite{mininni_rot_hel}. 

This sub-dominant direct energy cascade to the small scales provides in fact the needed small parameter for turbulence with rotation in the presence of helicity, in the form of the (adimensionalized) ratio $\chi=\epsilon/ L_F \tilde \epsilon$, where $\epsilon$ and $\tilde \epsilon$ are the energy and helicity fluxes, constant by definition in the inertial range. The energy flux to small scales is all the more negligible as more energy is transferred to large scales in an inverse cascade. The conjecture (now backed by numerical data) that $\chi \ll 1$ as the Rossby number gets small (but not necessarily tending itself to zero, see \cite{mininni_rot_hel}, the transition taking place for $Ro\approx 0.1$), led us in fact to the hypothesis that for moderate to strong rotation, the small-scale cascade is dominated by helicity and mediated by waves, leading to a prediction for the scaling $\zeta_2=1.5$ for rotating flows with maximum helicity \cite{mininni_rot_hel}. The lack of intermittency in the direct energy cascade is thus a novel feature providing for the first time a small parameter for small-scale three-dimensional turbulent flows, and giving a new way to look into turbulence.


{\it The authors would like to express their gratitude to J.R. Herring and J.J. Tribbia for their careful reading of the manuscript and comments, and to R. Rotunno and R. Wakimoto for useful discussions. Three-dimensional visualizations use VAPOR freeware \cite{Clyne07}. Computer time provided by NCAR, which is sponsored by NSF. PDM acknowledges support from the Carrera del Investigador Cient\'{\i}fico of CONICET.}

\end{document}